# Code2UML: Agentic LLMs with context engineering for scalable software visualization

Alin-Gabriel Văduva, Anca-Ioana Andreescu, Simona-Vasilica Oprea, Adela Bâra
Bucharest University of Economic Studies, Department of Economic Informatics and Cybernetics, no. 6 Piața Romană, Bucharest, Romania
*Corresponding author: alin.vaduva@csie.ase.ro

**Abstract:** Large Language Model (LLM)-based code analysis tools are adopted to automate software documentation tasks. However, the scalability of these approaches to real codebases, where Intermediate Representations (IR) exceed LLM context limits, remains underexplored. This paper introduces an agentic architecture with context engineering for automated UML diagram generation from source code repositories. It employs a hierarchy of five specialized agents: *PlannerAgent*, *AnalyzerAgent*, *DiagramAgent*, *CorrectorAgent* and *DependencyAnalyzerAgent*, built on the Claude Agent SDK, each addressing a distinct cognitive subtask. A deterministic, importance-weighted IR compaction layer transforms full project IRs into diagram-specific views guaranteed to fit within token constraints, requiring no LLM calls and completing in milliseconds. Thus, we evaluate the system across 12 open-source repositories in 4 programming languages (Java, JavaScript, PHP, Python) and 7 UML diagram types, producing 84 observations assessed on 5 automated metrics. Results demonstrate high syntactic validity (mean: 91.5%, with component and deployment diagrams reaching 100%), strong relationship precision (mean: 0.858) and consistent structural quality (mean: 81.7/100, with cross-language variance of 3.1 points). Entity recall averaged 0.313, reflecting deliberate architectural prioritization over exhaustive coverage. A sensitivity analysis (31 to 4,578 IR entities) confirms that quality scores remain stable regardless of scale.

## 1. Introduction

Software documentation remains one of the most neglected yet important aspects of software engineering practice. UML diagrams, when available and up-to-date, serve as the primary medium for communicating architectural decisions, behavioral flows and structural relationships among development teams [1]. However, the manual creation and maintenance of UML diagrams is time-consuming, error-prone and rapidly falls out of sync with evolving codebases. Studies consistently report that over 60% of software projects lack adequate architectural documentation and that existing diagrams become outdated within weeks of creation in actively developed projects [2], [3].

Advances in LLMs have opened new possibilities for automating software documentation tasks, including UML diagram generation from source code. However, applying LLMs to codebases presents a fundamental scalability challenge. The intermediate representations of medium-to-large projects (hundreds to thousands of classes, functions and relationships) routinely exceed the effective context window of even the largest available models [4], [5]. Naive approaches that attempt to pass entire codebase representations to a single LLM invocation result in truncated outputs, hallucinated relationships and diagrams that fail to capture architecturally significant structures. Furthermore, the diversity of UML diagram types, each with distinct notation rules, semantic constraints and information requirements, makes a one-size-fits-all generation approach insufficient [6], [7], [8].

Considering these challenges, our paper addresses the scalability and quality challenges of automated UML diagram generation by pursuing three primary research objectives. First, we design and implement a context engineering approach that deterministically transforms large code intermediate representations into compact, diagram-specific views guaranteed to fit within LLM context constraints, without requiring any LLM calls for the compaction itself. Second, we develop a multi-agent architecture where five specialized agents (*planner*, *analyzer*, *generator*, *corrector*, *dependencyanalyzer*) collaborate to produce syntactically valid PlantUML diagrams across 7 UML diagram types and 4 programming languages. Third, we empirically evaluate the system across 12 open-source repositories spanning two orders of magnitude in size (31 to 4,578 IR entities), assessing syntactic validity, entity recall, relationship precision, structural quality and complexity using five automated metrics.

The contributions of this work are threefold. First, we introduce a deterministic, importance-weighted IR compaction layer that transforms full project intermediate representations into diagram-specific views in milliseconds, requiring no LLM calls and no external services. This context engineering approach is complementary to prompt engineering: rather than optimizing instruction text, it optimizes the data payload that agents process, ensuring that architecturally significant elements survive even aggressive compression ratios. Unlike Retrieval-Augmented Generation (RAG) approaches that rely on embedding similarity to surface relevant code fragments, our compaction operates on structured intermediate representations with diagram-type-specific importance scoring, providing deterministic and reproducible view generation.

Second, we propose a hierarchy of five specialized agents built on the Claude Agent SDK, each addressing a distinct cognitive subtask. This multi-agent specialization replaces monolithic prompt-based approaches where a single LLM must simultaneously plan, analyze, generate and validate a design that mirrors established software engineering principles of separation of concerns and enables independent optimization of each component. The two-tier routing strategy (SINGLE path for architectural overviews, DEEP path for behavioral and structural diagrams) provides a framework for balancing generation cost against diagram fidelity. Third, we present a comprehensive empirical evaluation across 84 diagram-type observations (12 projects×7 diagram types).

The proposed solution is important because it addresses a critical limitation of current LLM-based software engineering tools. The inability to reliably process and document large-scale codebases within context window constraints. By combining deterministic context engineering with a hierarchy of specialized agents, the proposed architecture enables scalable, accurate and syntactically valid UML generation across multiple programming languages and diagram types. This contributes not only to improving automated software documentation and architectural understanding, but also to advancing the broader field of AI-assisted software engineering, where scalable and reliable agentic systems are increasingly needed for complex repository-level tasks.

## 2. Literature review

Early foundational research examined UML not only as a modeling language, but also as a formal reasoning framework. For example, [9] established that UML class diagrams can be formally analyzed through description logics, demonstrating both the computational complexity of reasoning over UML models and the feasibility of embedding reasoning capabilities into computer-aided software engineering tools. This work laid the theoretical groundwork for intelligent model validation and consistency checking. Extending model-driven transformations further, [10] combined transformer-based language understanding with a Domain-Specific Language (DSL) as an intermediate representation, enabling traceable conversion from textual requirements to UML models and ultimately to Python code.

Parallel to these developments, significant attention was directed toward generating software models directly from textual requirements. Early rule-based and NLP-driven systems attempted to extract classes, entities and relationships from natural language descriptions. For instance, [11] introduced a four-stage NLP framework involving preprocessing, sentence classification, syntactic parsing and rule-based UML generation, achieving strong classification performance. Similar efforts targeted Agile environments, where [12] proposed a semi-automated approach for generating use case diagrams from structured user stories using Stanford CoreNLP and handcrafted logical rules. Expanding Agile support, [13] integrated machine learning, SBERT embeddings and clustering methods to refine product backlogs, detect duplicate user stories and automatically generate UML class and use case diagrams for development teams.

The need to support multiple diagram perspectives motivated broader approaches to automated conceptual modeling. Addressing this challenge, [14] proposed a design science framework capable of generating four stylized UML diagram types: use case, class, activity and state machine diagrams from behavior-driven development scenarios. Expert evaluations confirmed the usefulness of such automatically generated models across requirements analysis, system design, implementation and testing. In related work, [15] focused on software product line engineering, proposing an automated method for generating feature models from UML use case and class diagrams while explicitly modeling data and actor perspectives. Similarly, [16] introduced a semi-automated methodology for generating goal models from UML use case and swimlane diagrams, separating behavioral, soft-goal, and constraint concerns to improve maintainability and evolution.

Beyond design modeling, UML has also been leveraged in software verification and testing. In this area, [17] proposed an automated model-based testing framework that transforms UML and IDEF models into executable test paths using graph-based representations. Experimental validation on an electricity billing system demonstrated strong test coverage and industrial applicability. Furthermore, [18] extended UML applications into project management by converting sequence diagrams into dependency graphs, Gantt charts and optimized scheduling plans using OCR, graph analysis and machine learning-based duration estimation.

Advances in LLMs have significantly transformed automated software modeling. Rather than relying solely on handcrafted rules, LLMs demonstrate stronger contextual understanding and reasoning over complex requirements. One notable study, [19], compared traditional NLP pipelines with a fine-tuned Qwen2.5-Coder model for class diagram generation, reporting substantial improvements in semantic fidelity, correctness and reduced manual correction effort. Similarly, [20] evaluated LLM-generated UML class diagrams against human solutions in an educational context, showing that although semantic errors remained challenging, LLM-generated diagrams achieved comparable syntactic and pragmatic quality. Complementing this work, [21] assessed the ability of LLMs to generate UML use case diagrams, finding performance levels similar to human participants in terms of completeness and redundancy.

LLM-based approaches have also been explored for behavioral modeling. For example, [22] investigated automatic generation of sequence diagrams from user stories, comparing a traditional automated pipeline with ChatGPT-generated outputs. Results indicated that conventional approaches produced simpler, requirement-aligned diagrams, while LLMs often introduced richer but occasionally over-specified interactions. Extending this integration further, [23] proposed a framework that combines LLM-generated UML behavioral models with formal logic engines, enabling automatic translation into logical specifications and deductive verification of software requirements.

At the same time, researchers have begun exploring multimodal applications of LLMs in software engineering. Addressing the challenge of digitizing informal design artifacts, [24] evaluated the ability of multiple LLMs to interpret hand-drawn UML class diagrams and convert them into formal machine-readable models. Other studies have focused on creating large-scale datasets and automated generation pipelines powered by LLMs. In [25], a dual-model framework combining LLaMA and DeepSeek was introduced for class diagram synthesis, producing a dataset of 5,000 samples containing technical specifications, PlantUML code and generated diagrams. To ensure structural correctness, the framework employed multimodal validation using multiple vision-language models. Building upon this architecture, [26] extended the approach to UML sequence diagrams, generating a benchmark dataset of 1,000 validated samples with strong semantic and structural consistency.

Beyond UML generation, LLM-based code understanding has also influenced software analysis. In the context of software security, [27] demonstrated that large-scale language models can successfully analyze source code, bytecode and even malformed code representations, achieving state-of-the-art defect detection performance in WebAssembly systems. Also, the emergence of autonomous agent systems has opened new directions for software engineering automation. A systematic review by [28] highlighted how large language model (LLM) agents can collaborate across software development lifecycle tasks, enabling scalable and autonomous software engineering workflows. Earlier work on intelligent multi-agent systems, such as [29], demonstrated the practical advantages of distributed agent coordination in real-world applications. More recently, [30] proposed a vision of software engineering agents equipped with memory, beliefs, intentions and normative reasoning, capable of collaborating with humans and other agents in software design, testing and deployment processes.

While substantial progress has been achieved in UML generation, validation and reasoning, challenges remain in ensuring semantic consistency, contextual adaptability and scalable collaboration between intelligent agents. These limitations motivate the development of more integrated, context-aware and autonomous systems for software modeling and engineering. A comparation is synthesized in Table 1.

Table 1. Comparison of the exiting studies

| Ref | Objective | Methods | Main Findings | Novelty / Contribution |
|---|---|---|---|---|
| [9] | Enable formal reasoning on UML class diagrams | Description logics, complexity analysis | Established EXPTIME complexity and decidable reasoning | Provided theoretical foundation for UML consistency checking |
| [10] | Convert textual specifications into UML and code | Transformer models+DSL+MDA pipeline | Improved traceability from text to executable code | Introduced DSL as semantic bridge between text and models |
| [11] | Generate UML class diagrams from textual requirements | NLP preprocessing, sentence classification, syntactic parsing, rule-based extraction | Achieved 88.46% classification accuracy and AUC of 0.9287 | Demonstrated effective rule-based NLP pipeline for requirement-to-UML transformation |
| [12] | Generate use case diagrams from user stories | Stanford CoreNLP, dependency parsing, logical rules | Improved extraction of actors and relationships | Addressed missing relationship modeling in use case generation |
| [13] | Improve Agile backlog management with UML automation | K-means clustering, SBERT embeddings, NLP extraction | Successfully identified duplicate stories and generated UML models | Integrated backlog refinement with automatic UML generation |
| [14] | Generate multiple UML models from user stories | NLP+BDD scenarios | Successfully produced four UML diagram types | Introduced multi-perspective UML generation |
| [15] | Generate feature models from requirements | UML use case+class diagram transformation | Successfully modeled features, actors, and data entities | Extended feature modeling with richer semantic entities |
| [16] | Generate goal models from UML requirements | Semi-automated extraction from use case and swimlane diagrams | Improved maintainability by separating goal concerns | Introduced multi-aspect goal modeling |
| [17] | Automate software testing from UML models | XMI parsing, graph construction, coverage analysis | Generated accurate test paths with high coverage | Linked UML models directly to test case generation |
| [18] | Transform sequence diagrams into project schedules | OCR, DAG construction, CPM, gradient optimization | Improved project scheduling accuracy | Connected UML design with project management |
| [19] | Compare NLP and fine-tuned LLMs for UML generation | Fine-tuned Qwen2.5-Coder model vs traditional NLP | LLM achieved 94% semantic fidelity and significantly reduced manual corrections | Showed superiority of domain-adapted LLMs over classical NLP approaches |
| [20] | Evaluate LLM agents for class diagram generation | Human vs LLM comparison on modeling exercises | LLMs achieved similar syntactic quality but weaker semantics | Highlighted strengths and limitations of LLM-generated UML |
| [21] | Evaluate LLM agents for use case diagram generation | Educational benchmark study | LLMs performed comparably to humans | Validated LLM utility in software education |
| [22] | Generate sequence diagrams from user stories | Rule-based pipeline vs ChatGPT | Traditional methods produced simpler, more aligned outputs | Compared symbolic and generative approaches |
| [23] | Integrate LLMs with formal verification | UML generation+logic translation+deduction | Improved requirements validation | Combined generative modeling with formal reasoning |
| [24] | Extract UML models from hand-drawn diagrams | Multimodal LLM image understanding | Generated usable formal models from sketches | Extended UML automation to visual inputs |
| [25] | Automate UML class diagram generation using LLMs | Dual-LLM pipeline (Meta LLaMA+DeepSeek DeepSeek), PlantUML generation, multimodal validation with VLMs | Generated 5,000 validated class diagram samples with strong structural consistency | Introduced large-scale benchmark dataset and multimodal validation for class diagram synthesis |
| [26] | Automate UML sequence diagram generation | Two-stage LLM pipeline with multimodal diagram validation | Produced 1,000 semantically aligned sequence diagrams | Extended multimodal UML synthesis from structural to behavioral diagrams |
| [27] | Analyze software artifacts using LLMs | Instruction embedding, source code embedding, LLM classification | Achieved >98% defect detection accuracy | Demonstrated LLM effectiveness beyond natural language artifacts |
| [28] | Review LLM-based multi-agent systems in software engineering | Systematic literature review + case studies | Identified strengths in agent collaboration and autonomy | Established research roadmap for multi-agent software engineering |
| [29] | Apply intelligent multi-agent systems in software applications | Distributed agent coordination | Improved system efficiency and resource allocation | Demonstrated practical value of agent collaboration |
| [30] | Define autonomous software engineering agents | BDI architecture, memory, normative reasoning | Proposed scalable human-AI collaboration model | Introduced cognitive agents for Software Engineering 2.0 |
| Our Work | Generate UML diagrams directly from source code repositories at scale | Multi-agent architecture (PlannerAgent, AnalyzerAgent, DiagramAgent, CorrectorAgent, DependencyAnalyzerAgent), deterministic IR compaction, context engineering, Claude Agent SDK | Achieved 91.5% syntax validity, 0.858 relationship precision, stable quality across 12 repositories, 4 languages, and 7 diagram types | First scalable agentic UML generation framework using deterministic context compression for repositories |

Our proposed framework differs from prior studies in three major ways: (a) Most prior work starts from natural language requirements or user stories, whereas our approach starts directly from source code repositories; (b) Previous LLM approaches rarely address large repositories exceeding context windows, whereas our work introduces deterministic IR compaction; (c) While [28]–[30] discuss multi-agent systems conceptually, our work provides an operational hierarchical agent architecture for UML generation.

## 3. Methodology

This section describes the agentic architecture for automated UML diagram generation from source code repositories, the context engineering strategies for managing large codebases within LLM context constraints and the evaluation framework. The architecture is built on the Claude Agent SDK with Claude Sonnet 4.6 as the underlying model, exposing a one-million-token context window alongside built-in filesystem tools (Read, Write, Glob, Grep). Figure 1 illustrates the complete pipeline.

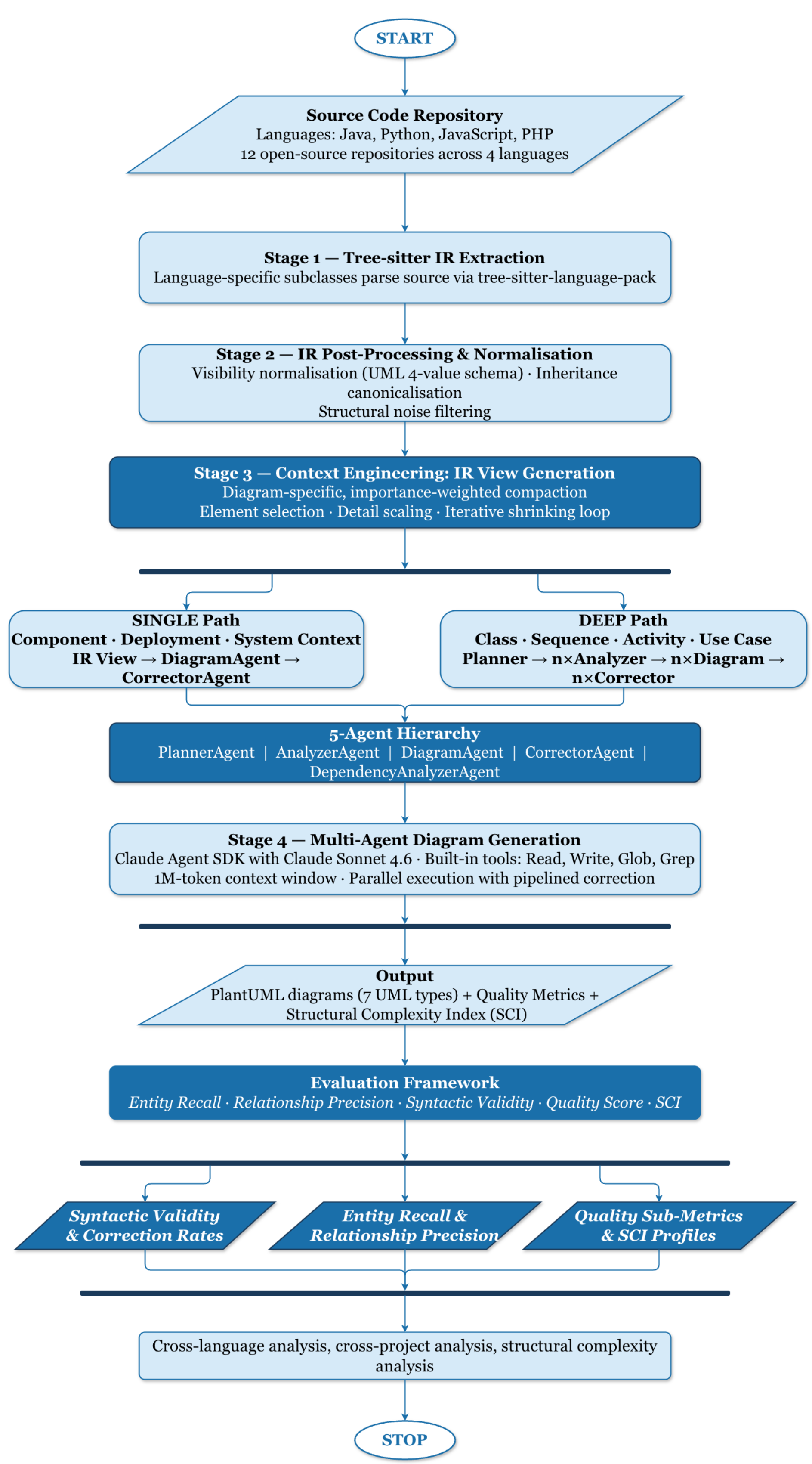

Figure 1. Research methodology pipeline

### 3.1. System architecture overview

The system follows a four-stage pipeline: (1) source code upload and project initialization, (2) IR generation via Tree-sitter parsing, (3) IR post-processing and normalization, and (4) multi-agent diagram generation driven by the Claude Agent SDK. The pipeline accepts codebases written in 4 programming languages (Java, Python, JavaScript and PHP) and produces 7 UML diagram types: class, sequence, activity, use case, component, deployment and system context. Language-specific tree-sitter extractors parse the uploaded source into a structured IR. A deterministic post-processing step normalizes language-specific constructs into a uniform schema shared across all downstream agents, ensuring that diagram agents operate identically regardless of source language.

### 3.2. IR and tree-sitter extraction

The IR is a language-agnostic JSON document encoding structural and relational properties of a source code repository. It contains collections for classes (with methods, attributes, inheritance), functions and metadata, decoupling parsing from diagram generation so that all downstream agents operate on the IR without re-parsing source code. Language detection selects the appropriate extractor from a registry. Mixed-language repositories run multiple extractors whose outputs merge into a single IR. Tree-sitter's error-tolerant parsing produces partial trees even for syntactically incomplete files, enabling recovery of structural information from imperfect code.

Following extraction, a post-processing step applies three classes of deterministic transformations to normalize the raw IR into the uniform schema consumed by all downstream agents: (1) visibility normalization, mapping language-specific access modifiers to the four-value UML schema (public, private, protected, package); (2) inheritance canonicalization, resolving fully qualified class names to simple names and removing duplicates; (3) structural noise filtering, removing empty or trivially small elements to improve the signal-to-noise ratio for downstream agents. The post-processed IR is used preferentially by all downstream agents.

### 3.3. Context engineering-IR view generation

Full IRs for medium-sized projects range from 200 KB to over 2 MB of JSON, far exceeding the effective token budget of a single agent invocation. The IR view generator addresses this through deterministic, importance-weighted compaction. A pure-Python transformation emits a diagram-specific view guaranteed to fall below a configurable byte threshold (60 KB for single-path, 100 KB for deep-path types). This requires no LLM calls and completes in milliseconds.

Compaction proceeds in three steps. First, element selection ranks all IR elements using a diagram-type-specific importance scoring function and retains the top candidates. For class diagrams, the importance score integrates method count, attribute count, presence of inheritance relationships and name-based heuristics prioritizing architecturally significant classes. For behavioral diagrams, function importance additionally weights functions containing call chains. Second, detail scaling adjusts per-element representation depth inversely with element count, retaining more methods and attributes for small projects and progressively fewer for large ones. Third, an iterative shrinking loop evaluates the estimated byte size after each compaction attempt and halves the element budget if the target is not met. Each diagram type receives a specialized view projection retaining only relevant IR fields. For example, the class view retains class definitions with methods, attributes and inheritance relationships, while the deployment view extracts infrastructure configuration metadata. The view file is written to the output directory before agent invocation.

### 3.4. Agent architecture-BaseAgent and the Claude agent SDK

All agents in the system are implemented as subclasses of an abstract *BaseAgent* class built on the Claude Agent SDK. The BaseAgent encapsulates the full agent session lifecycle. It initializes an agent session with a system prompt and a set of allowed tools, submits the task prompt and processes the resulting message stream. The message stream is translated into a normalized format consumed by the orchestrator. Reasoning text is extracted from assistant responses, tool invocations are collected as structured records and session-level statistics (duration, token usage breakdowns, cost) are forwarded to the *MetricsCollector* for computing efficiency metrics.

Each agent subclass is configured with a subset of filesystem tools appropriate to its role. The PlannerAgent, AnalyzerAgent, DiagramAgent and DependencyAnalyzerAgent use Read, Write, Glob and Grep, while the CorrectorAgent uses only Read and Write, since it operates on a single previously identified file and does not require filesystem discovery. The BaseAgent also exposes a hook mechanism through which the orchestrator registers callbacks for lifecycle events (before execution, after each message and on error), enabling integration with timing instrumentation and external monitoring systems.

### 3.5. Multi-agent hierarchy and orchestration

The DiagramOrchestrator class implements a two-tier routing strategy based on diagram type. Structural overview diagrams (component, deployment and system context) are routed through the SINGLE path, which invokes a single DiagramAgent with the pre-computed IR view as its primary information source. Behavioral and detailed structural diagrams (class, sequence, activity and use case) are routed through the DEEP path, which invokes a three-phase pipeline: PlannerAgent→parallel, AnalyzerAgents→parallel DiagramAgents with pipelined CorrectorAgents. After generation, all diagram types receive a correction pass from the CorrectorAgent regardless of path.

Five specialized agent subclasses implement the hierarchy. The PlannerAgent (Phase1 of the DEEP path) reads the compact IR view and produces a diagram plan listing the logical scopes (workflows, modules or functional areas) into which the project should be decomposed, scaled to project size by a rule table (one to three diagrams for projects with fewer than twenty classes, scaling up to fifteen to thirty diagrams for very large projects with more than two thousand classes). The AnalyzerAgent (Phase2) is instantiated once per planned scope and reads the specific source files identified by the planner, extracting a compact context containing actual method call chains, class relationships and flow logic for that scope. Multiple AnalyzerAgents execute in parallel to reduce latency. The DiagramAgent (Phase3) reads either the IR view (SINGLE path) or the enriched context (DEEP path) and generates a complete PlantUML file. DiagramAgents also execute concurrently in the DEEP path, with each CorrectorAgent launched immediately after its corresponding DiagramAgent completes (pipelined correction), so that correction of early-finishing diagrams overlaps with generation of later ones. The CorrectorAgent reads each generated *.puml* file, applies diagram-type-specific syntax validation rules and rewrites the file in place if corrections are needed. The DependencyAnalyzerAgent operates independently of the diagram generation pipeline and is invoked to analyze downloaded third-party library source code.

### 3.6. Agent system prompts and prompt engineering

Each agent subclass encodes its role and output requirements in a dedicated system prompt passed to the Claude Agent SDK at session initialization. The system prompts are the primary mechanism through which the system constrains each agent to a well-defined cognitive subtask and their design reflects both the structured nature of the generation problem and the idiosyncrasies of PlantUML syntax.

The DiagramAgent system prompts are organized around a shared base instruction block that is prepended to every diagram-type-specific section. The base block defines: (1) the agent's role as a software architecture expert; (2) the workflow for reading the IR view, identifying key elements, and generating PlantUML output; (3) universal PlantUML quality rules, including mandatory delimiters, prohibition of placeholder names, and a consistent styling template; and (4) a readability threshold limiting diagram size.

Each diagram type extends the base block with type-specific mappings that translate IR fields to PlantUML constructs (e.g., class inheritance hierarchies to UML generalization arrows, attribute types to composition relationships) and type-specific generation strategies. For instance, the sequence diagram prompt instructs agents to trace method call chains from source files, the activity diagram prompt emphasizes decision nodes and swim lane declarations, and the component and deployment prompts guide agents to discover infrastructure configuration files. The system context prompt provides a C4-style specification with stereotyped elements and a search protocol for identifying external integrations from framework annotations and infrastructure patterns.

The PlannerAgent prompt defines a structured output schema for scope decomposition, with a project-size-based scaling table governing diagram count. The AnalyzerAgent prompt encodes the context engineering principle of high-signal summarization: the agent is instructed to read source files but return only compact summaries capturing key participants, interaction flows, and relationships. The CorrectorAgent prompt is constructed dynamically by combining a base rule set with diagram-type-specific syntax validation rules that address known PlantUML syntax pitfalls for each diagram type. Beyond the system prompt, each DiagramAgent also receives a task-specific user prompt at invocation time that injects: (1) file system paths to the IR view and output directory; (2) scope parameters constraining the agent to a specific workflow or module in DEEP mode; (3) the path to the AnalyzerAgent's enriched context; (4) dependency context and documentation file paths when available.

### 3.7. Quality metrics framework

The evaluation framework defines five automated metrics computed after each diagram generation run by the MetricsCollector class. The metrics are designed to assess complementary dimensions of diagram quality: coverage, precision, structural richness, syntactic validity and generation efficiency.

Entity Recall measures the proportion of IR elements that appear in at least one generated diagram for that project and diagram type. It is computed as the count of diagram elements (extracted from the generated PlantUML by a regex-based parser) divided by the total number of elements in the Irm capturing the breadth of code coverage but does not distinguish between intentional abstraction (omitting utility classes from a class diagram) and unintentional omission.

$$Entity\ Recall\ =\frac{|Captured\ Entities|}{|IR\ Entities|} \tag{1}$$

Relationship Precision measures the fraction of generated relationships that connect entities declared in the same IR. It is computed per diagram as the count of valid relationships (those where both endpoints are declared in the IR or in other generated diagram elements) divided by the total number of generated relationships. A relationship precision of less than one indicates that the agent hallucinated one or more relationships involving entities not present in the source code.

$$Relationship\ Precision\ =\frac{|Valid\ Relationships|}{|Total\ Relationships|} \tag{2}$$

Syntactic Validity Rate is the percentage of generated *.puml* files that required no correction by the CorrectorAgent. This metric treats the CorrectorAgent as a reference validator. If the agent detects no syntax errors and makes no changes, the diagram is considered syntactically valid.

$$Validity\ Rate\ (\%)\ =\frac{Diagrams\ passing\ unchanged}{Total\ diagrams}\times 100 \tag{3}$$

The Quality Score is a composite metric computed from five sub-scores: Density Score (relationship-to-element ratio relative to UML best-practice ranges), Connectivity Score (proportion of elements with at least one connection), Labeling Score (proportion of relationships with meaningful labels), Documentation Score (proportion of elements with descriptive names or notes) and Structure Score (adherence to diagram-type-specific structural conventions). Each sub-score is normalized to a 0–100 scale and the Quality Score is their weighted average.

$$Q\ =\frac{1}{5}\left(S_{density}+S_{connect}+S_{labeling}+S_{document.}+S_{structure}\right) \tag{4}$$

The Structural Complexity Index (SCI) provides a size- and connectivity-adjusted measure of diagram richness, where $E$ is the element count and $R$ is the relationship count in a single diagram. It integrates the average node degree with a logarithmic penalization of density, ensuring that diagrams with many elements but few relationships score lower than diagrams with balanced element and relationship counts. The metric is grounded in graph theory (average degree as a connectivity measure) with diminishing returns at very high densities.

$$SCI\ =\ E\times\log_2\left(1+\frac{2R}{E}\right) \tag{5}$$

### 3.8. Evaluation corpus

The evaluation corpus comprises twelve open-source repositories across four programming languages: three Java projects (shopping-cart, Mindustry, debezium), three Python projects (GenericAgent, openai-agents, lyra), three JavaScript projects (evolver, impeccable, puter) and three PHP projects (laravel/ai, PHPMailer, composer). Projects were selected to represent a range of sizes, from small projects with as few as 31 IR elements (shopping-cart) to large projects with up to 4,578 IR elements (openai-agents), spanning nearly two orders of magnitude and a range of application domains including web APIs, game engines, data streaming platforms, autonomous agent frameworks, mailer libraries and dependency managers. For each project, seven diagram types were generated and evaluated, yielding 84 diagram-type-per-project observations. Each observation records the five primary metric values enumerated above plus element and relationship counts for the generated diagram. The Context Efficiency column (tokens per entity) was excluded from aggregate analysis as it requires live token counting instrumentation not available in the batch evaluation mode.

## 4. Results

This section presents the evaluation results across all twelve repositories and four programming languages. Results are reported both aggregated by diagram type (to reveal systematic differences in generation difficulty) and aggregated by language and project (to assess cross-language consistency and sensitivity to project size). Each subsection concludes with a visual analysis that highlights interaction effects and distributional patterns not readily apparent from the tabulated means alone.

### 4.1. Overall performance summary

Across all 84 observations, the system achieved a mean syntactic validity of 91.5%, a mean entity recall of 0.313, a mean relationship precision of 0.858 and a mean quality score of 81.7 out of 100. These results indicate that the multi-agent architecture with context engineering produces structurally sound, architecturally coherent diagrams for the majority of project-diagram type combinations. Table 2 summarizes the key metrics by diagram type.

Table 2. Mean performance metrics by diagram type across all 12 projects and 4 languages

| Diagram Type | Syntax Validity (%) | Entity Recall | Relationship Precision | Quality Score |
|---|---|---|---|---|
| ***Activity*** | 91.5 | 0.283 | N/A | 68.1 |
| ***Class*** | 96.3 | 0.445 | 0.917 | 83.6 |
| ***Component*** | 100.0 | 0.473 | 0.993 | 87.9 |
| ***Deployment*** | 100.0 | 0.181 | 0.815 | 78.0 |
| ***Sequence*** | 96.0 | 0.334 | 0.869 | 85.8 |
| ***System Context*** | 58.3 | N/A | 0.621 | 86.5 |
| ***Use Case*** | 98.6 | 0.165 | 0.933 | 81.7 |

As shown in Figure 2, the radar profiles reveal distinctive metric fingerprints for each diagram type. Class diagrams exhibit the broadest overall coverage, with strong performance across all five axes, whereas use case and deployment diagrams show characteristic indentations along the Entity Recall axis, consistent with their design intent of abstracting away source-level entities. Component diagrams achieve near-perfect Validity and Precision vertices, forming a distinctive peaked profile. The most pronounced inter-type variation occurs along the SCI axis, where activity diagrams extend to the perimeter while class and use case diagrams remain compressed, reflecting the wide range of structural complexity across diagram types.

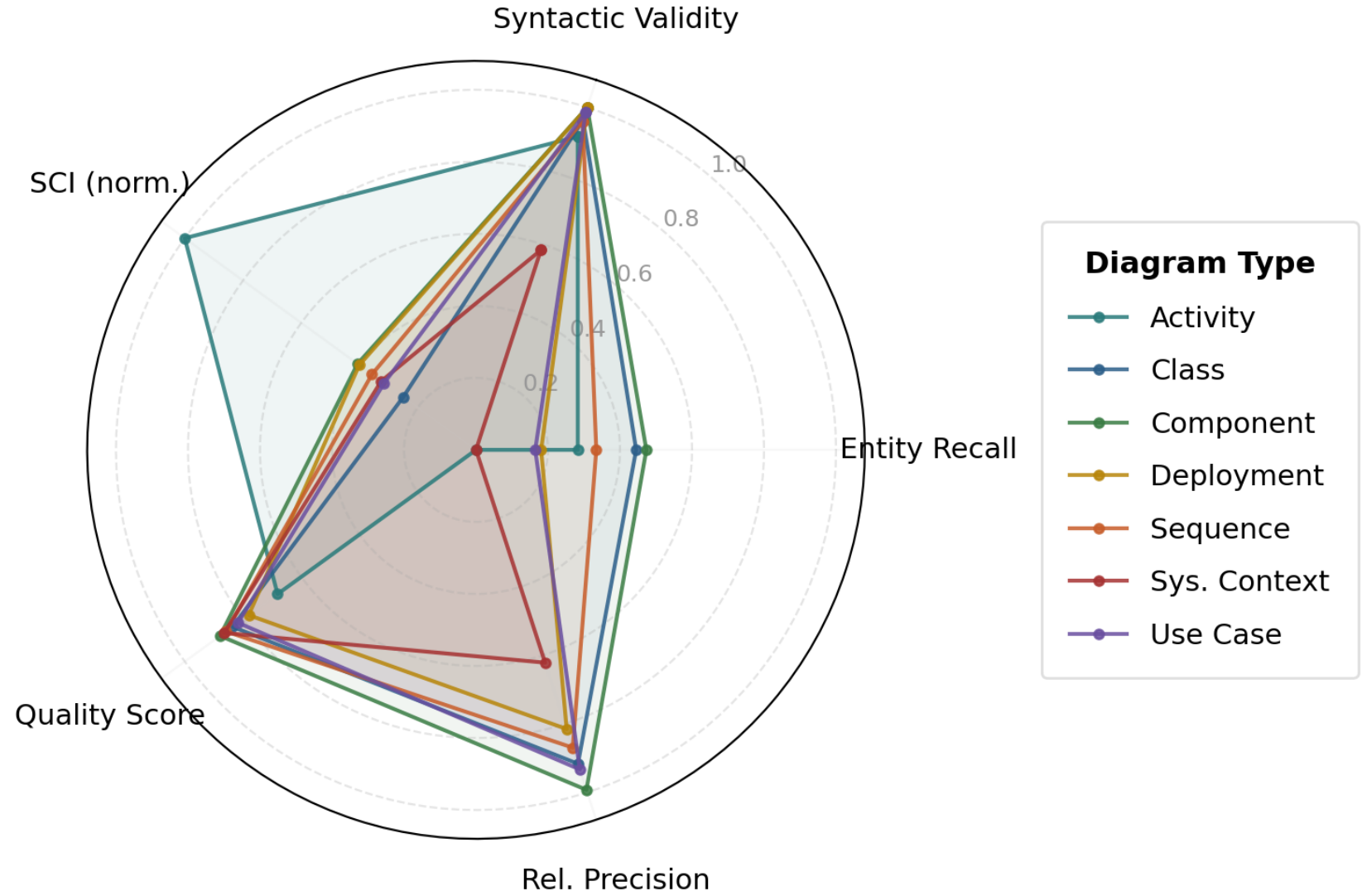

Figure 2. Radar chart of normalized metric profiles by diagram type

Figure 3 presents the quality scores remain remarkably stable across the full range of project sizes, with no systematic decline as IR entity counts increase from 60 to over 6,500 entities, confirming that the context engineering layer's adaptive compaction successfully maintains diagram quality regardless of project scale. Entity recall, by contrast, exhibits the expected inverse relationship with project size. The tallest bars appear at the left (small projects) and progressively shrink toward the right (large projects), reflecting the compaction algorithm's prioritization of architecturally significant elements at higher compression ratios. The interleaving of language colors across the sorted axis confirms that project size, not source language, is the dominant factor affecting recall.

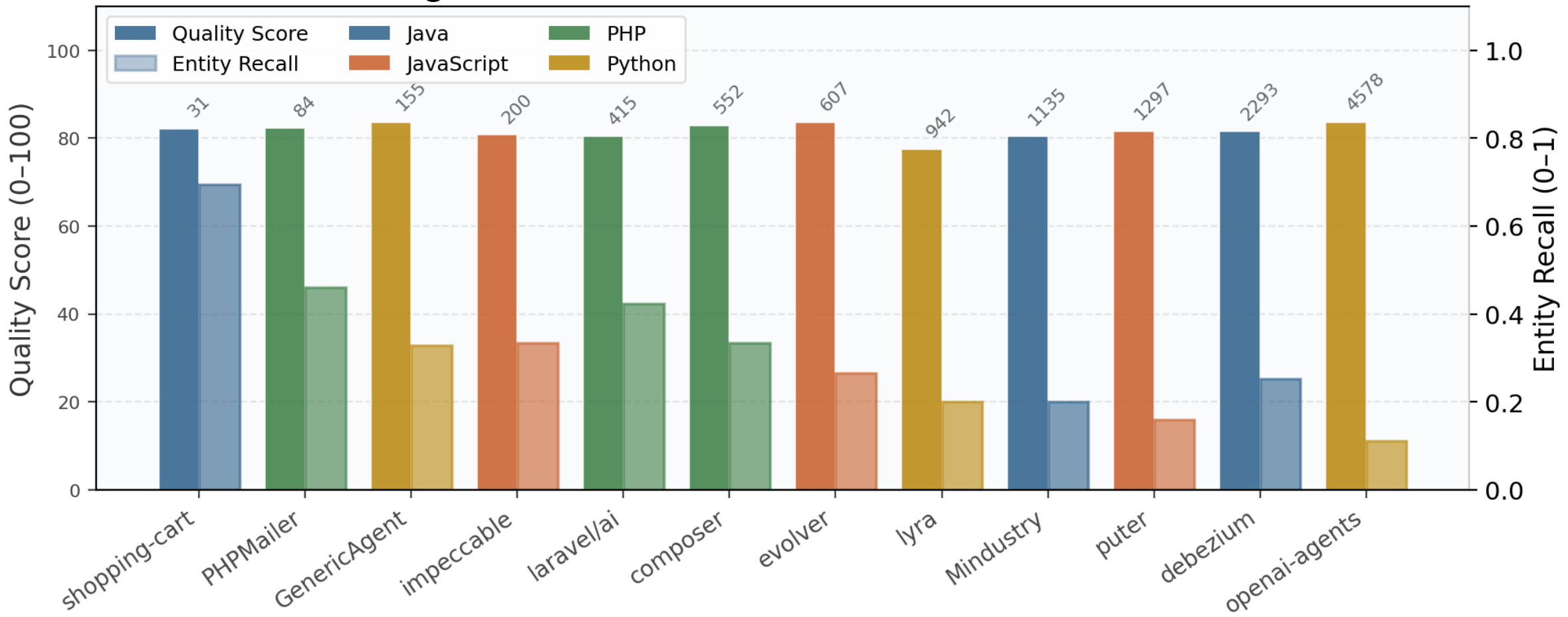

Figure 3. Quality score and entity recall per project (sorted by IR)

Figure 4 showcases the 4×7 interaction matrix, revealing that syntactic validity is predominantly determined by diagram type rather than by language. Columns exhibit far more color variation than rows. The Component and Deployment columns are uniformly deep green across all four languages, confirming their language-agnostic syntactic reliability. The System Context column contains the reddest cells, with some language–diagram combinations dropping to 0%, while others remain moderate, producing the highest inter-cell variance of any column. The row marginal means (μ values on the right) confirm the cross-language consistency observed in Table 3, with all four languages falling within a narrow band.

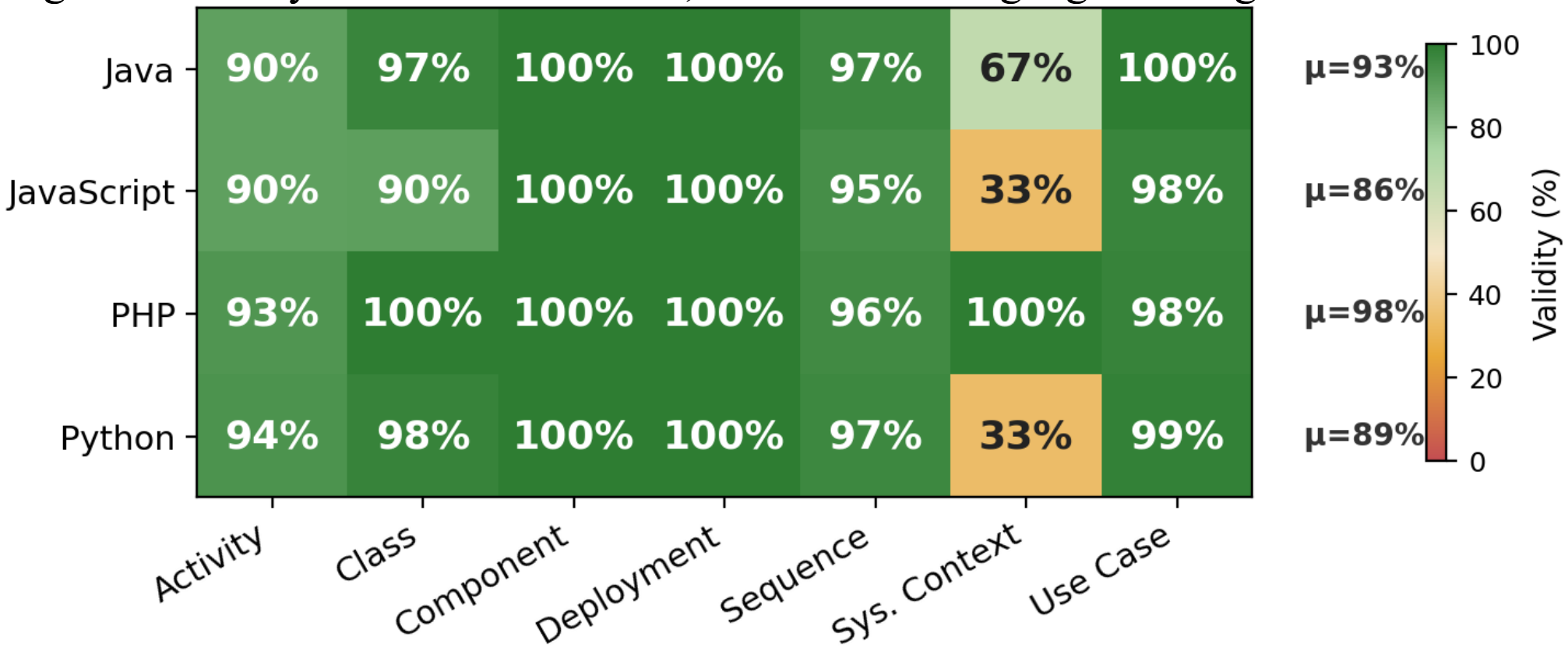

Figure 4. Heatmap of syntactic validity (%) across the full 4×7 language×diagram type interaction matrix

**4.2. Syntactic validity by diagram type**

Syntactic validity varies substantially by diagram type, reflecting differences in PlantUML syntax complexity and the effectiveness of the CorrectorAgent's type-specific rule sets. Component diagrams achieved the highest mean validity (100.0%) and deployment diagrams achieved 100.0%, both benefiting from simpler PlantUML syntax and well-established CorrectorAgent rules. Sequence diagrams achieved 96.0% validity, reflecting the effectiveness of the type-specific rules prohibiting invalid arrow modifiers and stereotypes in participant declarations. Activity diagrams achieved 91.5%, where the main failure mode was incorrect use of the continue keyword (not supported in PlantUML activity syntax) and elseif/else if confusion. Class diagrams achieved 96.3% and use case diagrams 98.6%.

System context diagrams achieves the lowest mean validity (58.3%), attributable to their use of C4-model-style PlantUML stereotypes and extended skinparam blocks, which push the limits of the current CorrectorAgent rule set. Several projects produced system context diagrams with 0% validity, indicating that the generated PlantUML contained structural errors that the corrector could not resolve with its current rule set. This finding motivates the extension of the CorrectorAgent with C4-specific validation rules as future work. Figure 5 disaggregates the syntactic validity and quality score results by both language and diagram type, revealing interaction effects that the type-level aggregates in Table 2 do not capture.

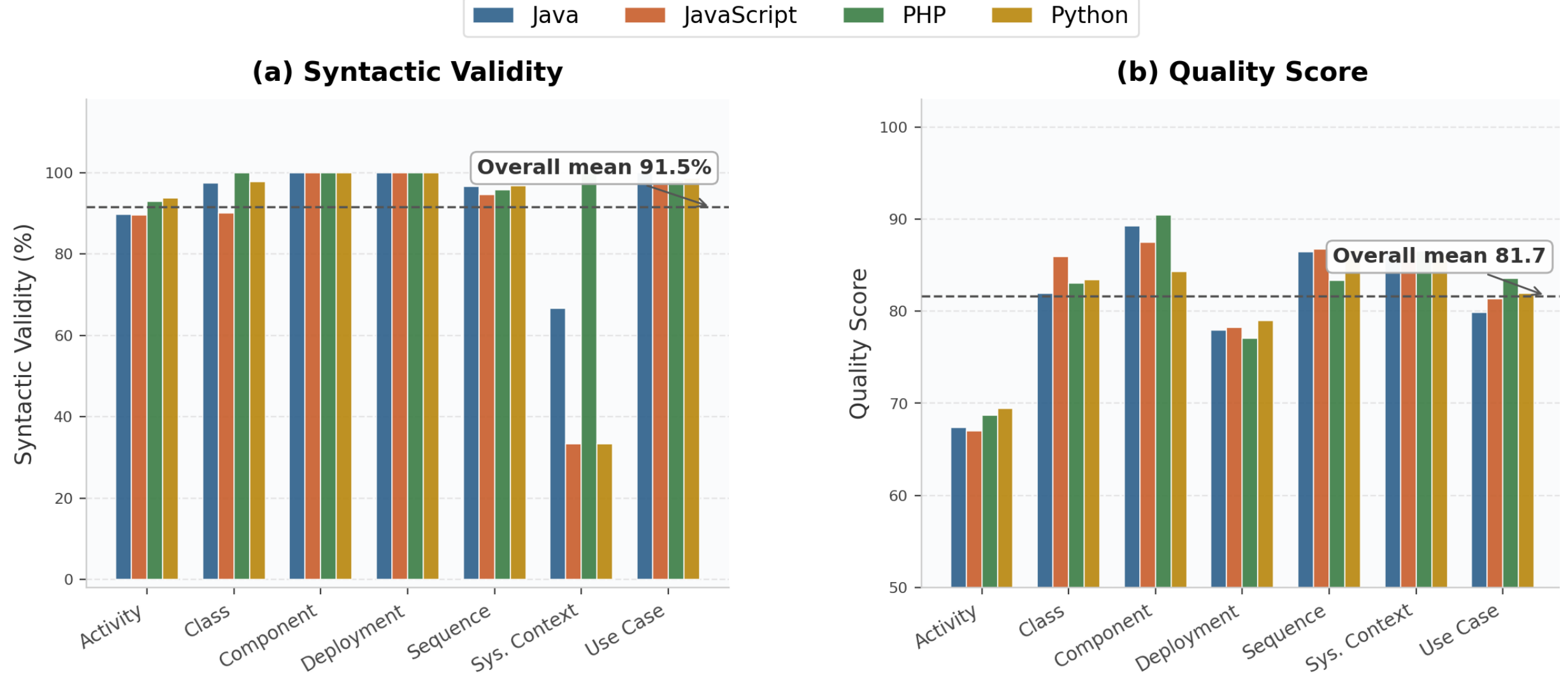

Figure 5. Mean syntactic validity (a) and quality score (b) by language and diagram type

As shown in Figure 5, the language×diagram interaction reveals patterns invisible in aggregated means. On the validity panel (a), the four language bars cluster tightly for component and deployment diagrams, indicating language-agnostic syntactic reliability. By contrast, system context diagrams exhibit the widest inter-language spread, with some languages achieving moderate validity while others collapse to zero, producing a bimodal distribution within that single diagram type. On the quality panel (b), language bars remain compressed across all seven types, with inter-language gaps rarely exceeding 5 points, confirming that the IR normalization layer absorbs language-specific structural variation before the diagram agents operate. A slight but consistent quality advantage for Java on behavioral diagram types (activity, sequence) reflects the richer type annotations available in Java IRs, which provide agents with more precise relationship targets.

**4.3. Entity recall and relationship precision**

Entity recall shows the expected pattern. Class diagrams achieved the highest recall (0.445), followed by component diagrams (0.473) and sequence diagrams (0.334). Deployment diagrams achieved low recall (0.181) because entity recall for deployment diagrams counts IR classes and functions against deployment-specific nodes (Docker containers, cloud services), which are architectural constructs not directly corresponding to source code entities. Use case diagrams achieved the lowest recall (0.165), consistent with their design intent of abstracting functional capabilities rather than enumerating code entities.

Relationship precision is consistently high across diagram types. Class diagrams achieved 0.917, sequence diagrams 0.869, component diagrams 0.993 and use case diagrams 0.933, indicating that the agents reliably generate relationships between entities that are actually declared in the source code, with minimal hallucination of spurious connections. The few cases of relationship precision below 1.0 were predominantly in sequence diagrams for large projects, where the agent occasionally generated message flows involving inferred intermediary components not explicitly declared in the IR.

**4.4. Quality sub-metric analysis**

Table 3 presents the five quality sub-metrics by diagram type. The quality scores decompose the aggregate quality measure into Density, Connectivity, Labeling, Documentation and Structure scores to reveal which aspects of diagram quality vary most across diagram types.

Table 3. Mean quality sub-metric scores by diagram type (scale: 0–100)

| Diagram Type | Quality Score | Density | Connectivity | Labeling | Documentation | Structure |
|---|---|---|---|---|---|---|
| ***Activity*** | 68.1 | 82.5 | 98.2 | 3.9 | 74.1 | 69.6 |
| ***Class*** | 83.6 | 99.3 | 64.6 | 90.2 | 71.4 | 88.8 |
| ***Component*** | 87.9 | 100.0 | 99.6 | 84.9 | 60.0 | 70.0 |
| ***Deployment*** | 78.0 | 99.3 | 73.2 | 98.9 | 60.0 | 11.7 |
| ***Sequence*** | 85.8 | 98.1 | 99.5 | 98.3 | 78.6 | 0.7 |
| ***System Context*** | 86.5 | 100.0 | 87.9 | 100.0 | 50.0 | 70.0 |
| ***Use Case*** | 81.7 | 99.6 | 95.5 | 52.1 | 69.8 | 70.5 |

Density and connectivity scores are uniformly high across diagram types (mean density: 97.0, mean connectivity: 88.4), indicating that the agents reliably generate well-proportioned, fully connected diagrams regardless of type. Labeling scores (mean: 75.5, std: 35.2) and structure scores (mean: 54.5, std: 32.9) exhibit the highest inter-diagram variance, reflecting inherent differences in UML notation conventions. Activity diagrams have near-zero labeling scores because activity flow connections are unlabeled by convention, while component and system context diagrams achieve high labeling scores because their dependency arrows carry protocol and data flow labels. Documentation scores (mean: 66.3) are moderately high across all diagram types, with sequence and class diagrams achieving the highest values due to the richness of method-level naming information available in their respective IR projections.

Figure 6 visualizes the sub-metric composition as stacked horizontal bars, providing a profile view of how each diagram type achieves its composite quality score.

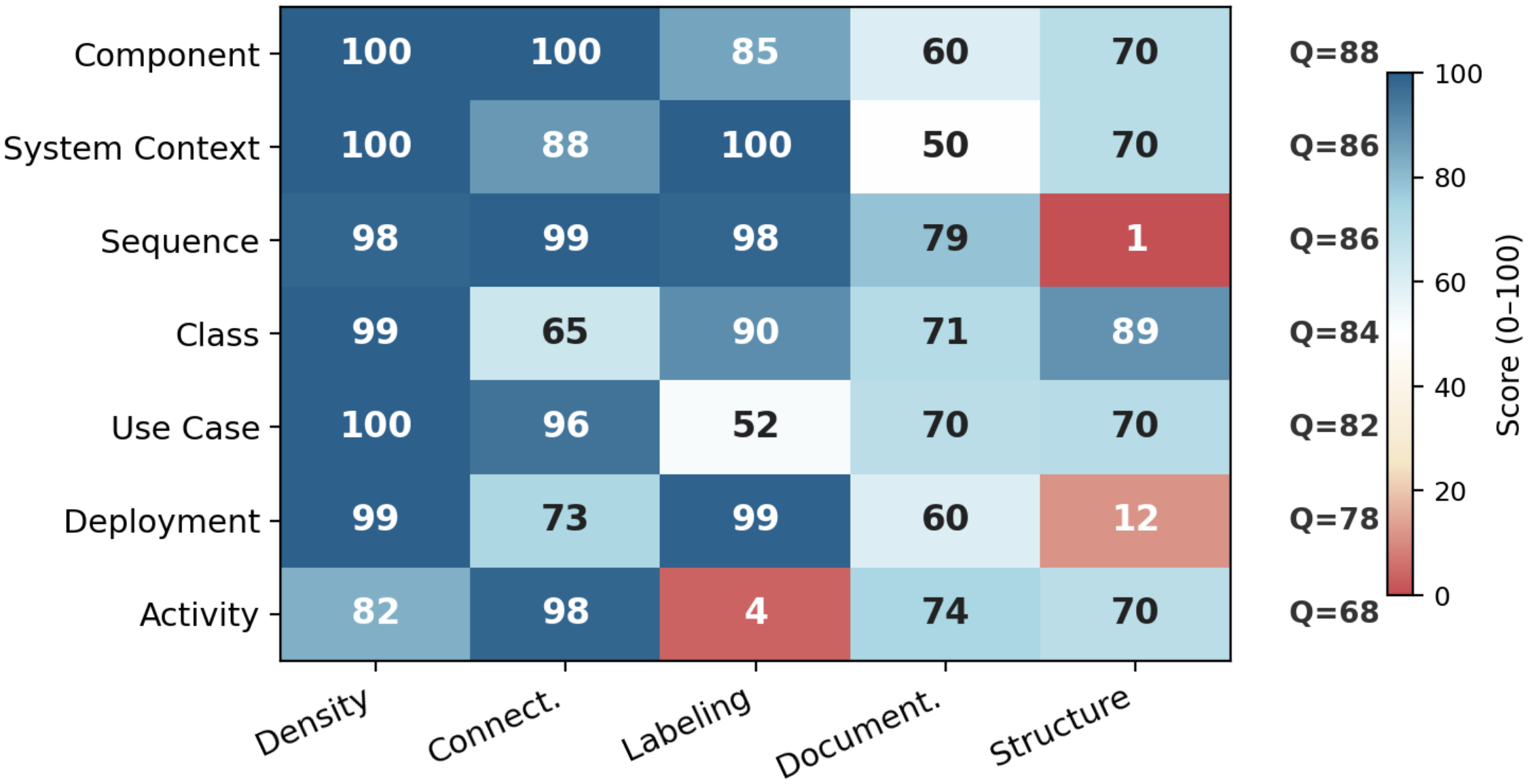

Figure 6. Heatmap of quality sub-metric scores (0–100) by diagram type, sorted by composite Quality Score (Q). Darker blue indicates higher scores, darker red indicates lower scores.

As shown in Figure 6, the heatmap reveals sharply distinct quality fingerprints across diagram types. The most striking pattern is the Activity row, where the Labeling cell drops to near zero while Density and Connectivity remain saturated, confirming that activity flow edges are inherently unlabeled by UML convention. Component diagrams display the most uniform row coloring, indicating that the SINGLE-path agent produces well-rounded diagrams for this type. Comparing the Sequence and Deployment rows reveals an instructive contrast. Both achieve similar composite Quality Scores, yet Sequence derives its strength from Documentation and Connectivity (reflecting rich participant naming and dense message flows), while Deployment relies on Structure and Labeling (reflecting correctly typed infrastructure nodes and annotated relationships).

### 4.5. Cross-language and cross-project analysis

Table 4 presents the mean performance metrics aggregated by programming language across all diagram types and projects.

Table 4. Mean performance metrics (aggregated across all 7 diagram types and 3 projects per language)

| Language | Syntax Validity (%) | Entity Recall | Relationship Precision | Quality Score |
|---|---|---|---|---|
| ***Java*** | 92.9 | 0.382 | 0.853 | 81.3 |
| ***JavaScript*** | 86.4 | 0.253 | 0.801 | 81.9 |
| ***PHP*** | 98.1 | 0.406 | 0.939 | 81.8 |
| ***Python*** | 88.6 | 0.213 | 0.840 | 81.6 |

Cross-language consistency is  high. Quality scores range from 81.3 to 81.9 across languages, a range of only 0.6 points. Syntactic validity is similarly consistent, ranging from 86.4% to 98.1% across languages. These narrow cross-language ranges demonstrate that the IR normalization and context engineering layers effectively abstract away language-specific structural differences, enabling the diagram agents to operate uniformly on the standardized IR representation regardless of the source language.

Table 5 disaggregates results by individual project to examine sensitivity to project size. IR entity counts range from 31 elements (Java/shopping-cart) to 4578 elements (Python/openai-agents), spanning nearly two orders of magnitude. Despite this wide range, quality scores are stable across project sizes.

Table 5. Performance metrics by individual project (sorted by language and project ID)

| Language | Project | IR Entities (avg) | Syntax Validity (%) | Entity Recall | Quality Score |
|---|---|---|---|---|---|
| ***Java*** | ***shopping-cart*** | 31 | 85.7 | 0.695 | 82.1 |
| ***Java*** | ***Mindustry*** | 1135 | 96.9 | 0.200 | 80.4 |
| ***Java*** | ***debezium*** | 2293 | 96.2 | 0.252 | 81.5 |
| ***JavaScript*** | ***evolver*** | 607 | 83.2 | 0.265 | 83.5 |
| ***JavaScript*** | ***impeccable*** | 200 | 80.1 | 0.334 | 80.7 |
| ***JavaScript*** | ***Puter*** | 1297 | 96.0 | 0.159 | 81.5 |
| ***PHP*** | ***laravel/ai*** | 415 | 96.6 | 0.423 | 80.4 |
| ***PHP*** | ***PHPMailer*** | 84 | 98.6 | 0.460 | 82.2 |
| ***PHP*** | ***composer*** | 552 | 99.2 | 0.334 | 82.8 |
| ***Python*** | ***GenericAgent*** | 155 | 84.3 | 0.328 | 83.6 |
| ***Python*** | ***openai-agents*** | 4578 | 82.7 | 0.110 | 83.6 |
| ***Python*** | ***Lyra*** | 942 | 98.7 | 0.201 | 77.5 |

Quality scores across projects range from 77.5 to 83.6, with no systematic ordering by IR entity count, confirming that the context engineering layer's adaptive detail scaling successfully maintains diagram quality across the full spectrum of project scales evaluated. Entity recall shows the expected inverse relationship with project size. Smaller projects (fewer than 100 IR entities) achieved higher mean recall (0.578) than larger projects (more than 500 IR entities, mean recall: 0.217), consistent with the compaction algorithm's prioritization of the most important elements at higher compression ratios.

Figure 7 plots entity recall against IR entity count for every project–diagram type observation, color-coded by language and shaped by diagram type, to expose the distributional relationship between project scale and coverage.

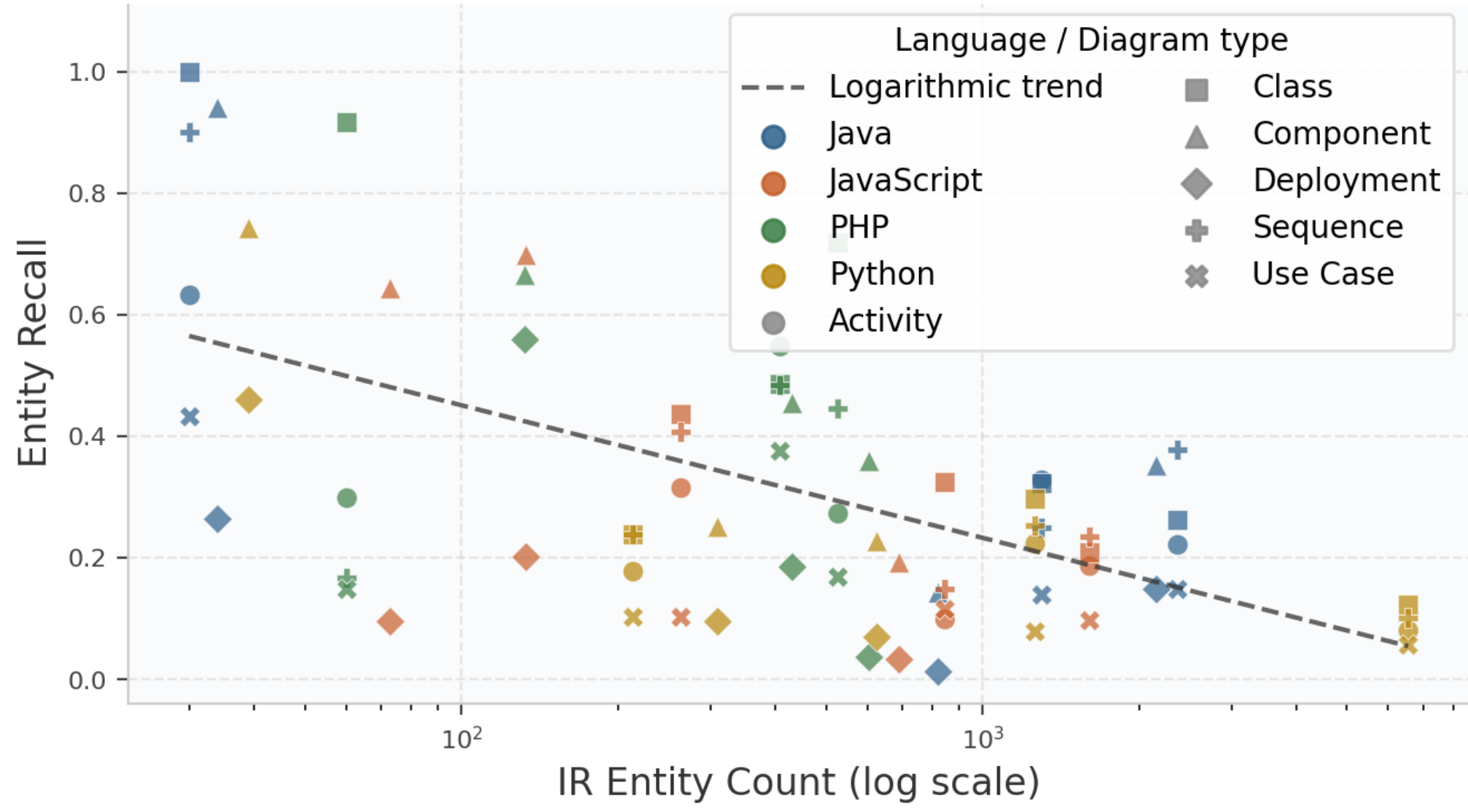

Figure 7. Entity recall vs. IR entity count (log scale)

The scatter exhibits a clear logarithmic decay envelope. Recall decreases steeply as IR size rises from 60 to approximately 500 entities, then flattens into a plateau between 0.05 and 0.20 for projects exceeding 1,000 entities. This two-regime shape reflects the compaction algorithm's iterative halving of max_elements. Small projects undergo little or no compaction (retaining most entities), while large projects trigger multiple shrinking iterations that converge to a stable, architecturally focused subset. Notably, the four language colors intermingle throughout the scatter without forming language-specific clusters, confirming that project size, not source language, is the dominant predictor of recall. Class diagram markers consistently sit above the trend line at every project size, reflecting their direct mapping from IR classes to diagram elements, while use case and deployment markers cluster below the trend, confirming the metric mismatch discussed above.

### 4.6. Diagram structural complexity

The SCI provides a size- and connectivity-adjusted measure of diagram richness. Table 6 presents the mean SCI values along with element and relationship counts by diagram type.

Table 6. Mean diagram elements, relationships and SCI by diagram type.

| Diagram Type | Mean Elements | Mean Relationships | Mean SCI | SCI Std |
|---|---|---|---|---|
| ***Activity*** | 54.4 | 64.3 | 95.1 | 19.2 |
| ***Class*** | 19.7 | 13.0 | 23.8 | 4.2 |
| ***Component*** | 16.2 | 38.9 | 38.7 | 16.5 |
| ***Deployment*** | 29.7 | 21.5 | 38.1 | 13.2 |
| ***Sequence*** | 9.5 | 52.6 | 34.0 | 5.8 |
| ***System Context*** | 20.7 | 19.2 | 30.9 | 7.8 |
| ***Use Case*** | 17.7 | 19.9 | 30.0 | 6.1 |

Activity diagrams had the highest SCI (95.1, std=19.2) driven by large element counts and dense control flows. Sequence diagrams, despite having the fewest elements (9.5), achieved moderate SCI (34.0) due to their high relationship-to-element ratio (52.6 messages). Class diagrams showed the lowest SCI (23.8), reflecting consistent compact sub-diagrams from DEEP-path scope decomposition. Figure 8 visualizes these structural profiles. It presents the structural profile of each diagram type as a lollipop chart, plotting the three constituent metrics (elements, relationships and SCI) on a common horizontal axis to reveal how these quantities relate within and across types.

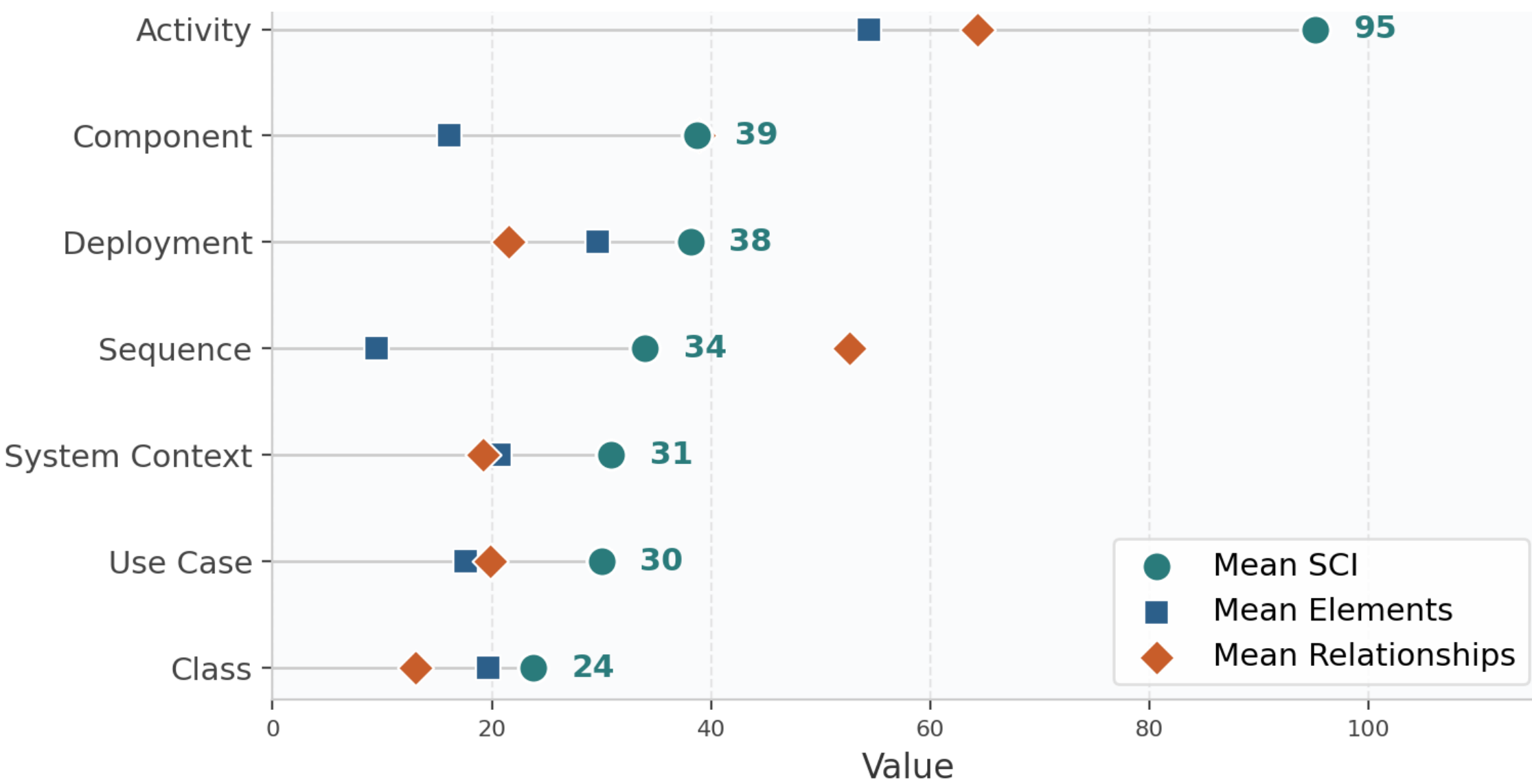

Figure 8. Mean elements (squares), relationships (diamonds), and SCI (circles, lollipop) per diagram type, sorted by SCI

As shown in Figure 8, the lollipop chart exposes structural asymmetries that the tabulated means do not convey. Sequence diagrams are a clear outlier. Their relationship diamond sits far to the right of their element square, producing the widest horizontal gap of any diagram type, quantifying the exceptionally high message-to-participant ratio characteristic of sequence diagrams. Activity diagrams show both markers positioned far right and close together, indicating that their high SCI is driven by sheer scale (many nodes and many edges) rather than by disproportionate connectivity. Class diagrams cluster at the left end with tightly grouped markers, reflecting compact, structurally uniform sub-diagrams, a pattern consistent with the DEEP path's scope decomposition, which partitions large class hierarchies into focused sub-diagrams. Component and deployment diagrams occupy a balanced middle band where elements and relationships are nearly equal, mirroring the one-to-one mapping between architectural components and their inter-component dependencies.

### 4.7. Ablation study

We simulate the removal of the CorrectorAgent through a post-hoc decomposition of the evaluation results. Since the CorrectorAgent operates as a pure post-processing step that does not alter diagram content (entity recall, relationship precision and quality scores remain unchanged regardless of correction), its contribution can be isolated by examining the proportion of diagrams that achieved syntactic validity without requiring any corrector intervention. This simulation provides a lower-bound estimate of the CorrectorAgent's impact. The results are reported in Table 7.

Table 7. Syntactic validity excluding the CorrectorAgent

| Diagram Type | No Correction Needed | Partially Corrected | Uncorrectable | Mean Validity (%) |
|---|---|---|---|---|
| ***Activity*** | 2/12 (16.7%) | 10/12 (83.3%) | 0 | 91.5 |
| ***Class*** | 7/12 (58.3%) | 5/12 (41.7%) | 0 | 96.3 |
| ***Component*** | 12/12 (100%) | 0 | 0 | 100.0 |
| ***Deployment*** | 12/12 (100%) | 0 | 0 | 100.0 |
| ***Sequence*** | 6/12 (50.0%) | 6/12 (50.0%) | 0 | 96.0 |
| ***System Context*** | 7/12 (58.3%) | 0 | 5/12 (41.7%) | 58.3 |
| ***Use Case*** | 9/12 (75.0%) | 3/12 (25.0%) | 0 | 98.6 |
| ***Overall*** | 55/84 (65.5%) | 24/84 (28.6%) | 5/84 (6.0%) | 91.5 |

To assess the contribution of the CorrectorAgent, we perform a post-hoc ablation by decomposing the evaluation results based on correction dependency. Since the CorrectorAgent operates as a pure post-processing step that does not alter diagram content (entity recall, relationship precision and quality scores remain unchanged regardless of correction), its impact can be isolated by examining which diagram batches required syntactic intervention. Only 65.5% of observations (55/84) produced syntactically valid diagrams without any corrector intervention. The remaining 34.5% either required partial correction (28.6%) or contained errors beyond the corrector's rule set (6.0%). The ablation reveals three distinct correction regimes: (1) Component and Deployment diagrams never require correction, confirming that the SINGLE-path DiagramAgent reliably produces valid PlantUML for simpler syntactic forms; (2) Activity diagrams exhibit the highest correction dependency (83.3% of batches required intervention), reflecting the complexity of PlantUML's activity syntax with swim lanes, decision nodes and prohibited keywords; (3) System Context diagrams represent a failure mode where 41.7% of outputs contain C4-model stereotypes that fall entirely outside the CorrectorAgent's current rule set, rendering them uncorrectable. These findings confirm that the CorrectorAgent is architecturally essential for behavioral diagram types but motivate rule set extension for C4-specific constructs.

## 5. Discussion

The results demonstrate that agentic architectures with context engineering produce structurally sound UML diagrams from source code across multiple programming languages. The IR view generation layer proved essential to system performance. Without it, large project IRs (exceeding 500 KB) would overwhelm the agent's effective context window, producing either truncated or hallucinated diagrams. The importance-weighted compaction ensures that architecturally significant elements (services, controllers, orchestrators, domain models) are preserved even at aggressive compression ratios, as evidenced by the stability of quality scores across project sizes (range: 77.5–83.6). The iterative shrinking algorithm guarantees that the view fits within the SDK's Read tool constraints while maintaining element diversity. This approach represents a form of context engineering that is complementary to prompt engineering. Rather than optimizing the instruction text, it optimizes the data payload that the agent processes.

The two-tier orchestration model, single mode for architectural diagrams and deep analysis for behavioral and structural diagrams, reflects a deliberate trade-off between generation cost and diagram fidelity. Single-mode diagrams (component, deployment, system context) achieved the highest syntactic validity (100% for the first two types) with a single agent invocation, confirming that high-level architectural views can be reliably derived from the IR view alone. Deep-analysis diagrams (class, sequence, activity, use case) invest three to four agent invocations per scope (planner, analyzer, generator, corrector) but produce source-code-grounded diagrams with higher entity recall (class: 0.445, sequence: 0.334 vs. deployment: 0.181). The parallel execution of AnalyzerAgents and DiagramAgents, combined with pipelined correction, mitigates the latency overhead inherent in the multi-agent approach.

The CorrectorAgent plays a critical role in achieving high syntactic validity rates. Without correction, raw LLM-generated PlantUML frequently contains subtle syntax errors: unbalanced braces, invalid keywords (e.g., device instead of node) and forbidden constructs (e.g., skinparam linetype ortho, continue in activity diagrams). The type-specific rule sets encoded in the corrector's system prompt enable targeted fixes without altering diagram semantics. The contrast between 100% validity for component and deployment diagrams versus the lower validity for system context diagrams suggests that diagram types with simpler PlantUML syntax benefit most from the current rule set, while C4-style stereotyped syntax requires rule set extension. The moderate overall entity recall (0.313) should be interpreted in the context of diagram purpose. UML diagrams are not intended to exhaustively enumerate all code entities but rather to provide focused views of architecturally significant structures. A class diagram covering 30–45% of IR entities may still be a high-quality diagram if it captures the core domain model and key relationships, as reflected in the consistently high quality scores (mean: 81.7).

The separation of concerns across the agent hierarchy mirrors established software engineering principles. The PlannerAgent's role as a project decomposition reasoner, the AnalyzerAgent's role as a high-signal code summarizer, the DiagramAgent's role as a PlantUML generator and the CorrectorAgent's role as a syntax validator each address a distinct cognitive subtask, enabling each agent's system prompt to be optimized for a specific, well-defined task rather than requiring a single monolithic prompt to handle planning, analysis, generation and correction simultaneously. The result is a system where each component is individually tunable and where failures are localized. A CorrectorAgent failure affects only syntax correction, not the underlying diagram structure.

## 6. Conclusions

This paper presented an agentic architecture for automated UML diagram generation from source code repositories, built on the Claude Agent SDK with Claude Sonnet 4.6 as the underlying LLM. The multi-agent architecture streamlines the diagram generation process by operating directly on structured intermediate representations without requiring vector databases, embedding models or complex graph-based workflow orchestration. The core contribution is the integration of a context engineering layer, comprising diagram-specific IR view generators with importance-weighted compaction, with a hierarchy of five specialized agents (PlannerAgent, AnalyzerAgent, DiagramAgent, CorrectorAgent and DependencyAnalyzerAgent) that collectively transform structured intermediate representations into syntactically valid, architecturally focused UML diagrams.

The context engineering approach represents a methodological contribution that addresses a fundamental challenge in applying LLMs to software engineering tasks: managing the mismatch between the scale of codebases and the effective context window of LLM-based agents. Rather than relying on retrieval-based approaches to surface relevant code fragments, the system performs deterministic, importance-weighted compaction of the full intermediate representation into diagram-specific views. This pure-Python transformation requires no LLM calls, no external services and completes in milliseconds, producing a compact payload (under 60 KB for single-diagram types, under 100 KB for deep-analysis types) that fits within the Claude Agent SDK's Read tool constraints. The iterative shrinking algorithm ensures that the compaction adapts to project size. Small projects retain full detail (up to 20 methods and 20 attributes per class), while large projects are progressively compressed (down to 5 methods and 5 attributes) until the byte-size target is met, with importance scoring ensuring that architecturally significant entities survive the compaction.

The empirical evaluation across 12 open-source repositories in four programming languages (Java, JavaScript, PHP, Python) produced 84 diagram-type observations assessed on five automated metrics. The results demonstrate that the agentic architecture achieves high syntactic validity (overall mean: 91.5%, with component and deployment diagrams reaching 100%), strong relationship precision (mean: 0.858, indicating that the vast majority of generated relationships connect valid, declared entities) and consistent structural quality (mean: 81.7 out of 100, with cross-language variance of only 0.6 points). Entity recall averaged 0.313 across all diagram types and languages, with class diagrams achieving the highest recall (0.445) and use case diagrams the lowest (0.165). The context engineering layer prioritizes architectural significance over exhaustive coverage, producing focused diagrams rather than comprehensive but unwieldy entity enumerations.

The methodological significance of this work lies in demonstrating that multi-agent specialization combined with context engineering outperforms monolithic prompt-based approaches for complex software engineering tasks. Each agent in the hierarchy addresses a distinct cognitive subtask. The parallel execution model (concurrent AnalyzerAgents, concurrent DiagramAgents with pipelined correction) demonstrates that agentic systems can be engineered for throughput as well as quality, a practical requirement for processing repositories with hundreds of classes across multiple diagram types.

Several limitations constrain the generalizability of these findings. The automated evaluation metrics, while reproducible and scalable, do not capture the subjective usefulness of diagrams to human software engineers. A diagram with low entity recall may nonetheless be highly informative if it highlights the correct architectural patterns. The system context diagrams' low syntactic validity indicates that the CorrectorAgent's rule set requires extension to handle C4-model-specific PlantUML constructs. Additionally, while the evaluation covered four typologically diverse programming languages, extending to additional languages and to industrial codebases would strengthen the generalizability claims.

Future work should pursue several directions. First, adaptive context engineering strategies that dynamically adjust compaction parameters based on diagram quality feedback could improve the trade-off between coverage and quality.Second, extending the CorrectorAgent with specialized C4-model syntax validation rules would address the system context diagram validity gap. Third, exploring multi-turn agent interactions where the DiagramAgent iteratively refines its output based on automated validation feedback could improve both syntactic validity and structural quality without additional human oversight. Finally, investigating the transferability of the context engineering approach to other software engineering tasks (automated documentation generation, code review, test case generation) would establish the generality of the importance-weighted compaction methodology beyond UML diagram generation.

**Acknowledgement**. This work was supported by a grant of the Ministry of Research, Innovation and Digitization, CNCS/CCCDI - UEFISCDI, project number COFUND-CETP-SMART-LEM-1, within PNCDI IV. This research was funded by CETP, the Clean Energy Transition Partnership under the 2022 CETP joint call for research proposals, co funded by the European Commission (GAN° 101069750) and with the funding organizations detailed on https://cetpartnership.eu/funding-agencies-and-call-modules.
**Disclosure statement (Competing interest)**. The authors have no relevant financial or non-financial interests to disclose.
**Data availability statement**. Details are provided in a public repository: https://github.com/alingabriel743/uml-study-data
**Ethical approval**. Not applicable.
**Informed consent**. Not applicable.
**Author contributions**. **AV**: Conceptualization, Methodology, Software, Investigation, Resources, Data Curation, Writing-Original Draft, Validation, Formal analysis. **AB, AA**: Formal analysis, Investigation, Writing-Original Draft, Writing-Review and Editing, Supervision; **SVO**: Formal analysis, Investigation, Writing-Original Draft, Writing-Review and Editing, Visualization, Validation, Project administration, Supervision.